\documentclass[sigconf,10pt]{acmart}
\pdfoutput=1
\usepackage{array}
\usepackage{datetime}
\usepackage{url}
\usepackage{verbatim} 
\usepackage{xspace} 
\usepackage{amsthm}
\usepackage{framed}
\usepackage{changepage}
\usepackage{multirow}
\usepackage{graphicx}
\usepackage{caption}
\usepackage{scalerel}
\usepackage{xcolor}
\usepackage[
  separate-uncertainty = true,
  multi-part-units = repeat
]{siunitx}

\usepackage[
 font=large 
 ]{subfig}
 
\usepackage{titlesec}
\titlespacing*{\subsection}{5pt}{0.5\baselineskip}{2pt}

\usepackage{soul}
\usepackage[normalem]{ulem}

\setstcolor{red}
\newcommand\redsout{\bgroup\markoverwith{\textcolor{red}{\rule[0.5ex]{2pt}{1.0pt}}}\ULon}
\newcommand{\errata}[1]{#1}
\newcommand{\sklee}[1]{#1}

\definecolor{dgreen}{HTML}{228B22}

\newcommand\blfootnote[1]{%
  \begingroup
  \renewcommand\thefootnote{}\footnote{#1}%
  \addtocounter{footnote}{-1}%
  \endgroup
}

\newcommand{\mycaption}[2]{\caption{\textbf{#1}. {#2}}}
\newcommand{\sref}[1]{\S\ref{#1}}
\newcommand{\vheading}[1]{\vspace{0.05in}\noindent\textbf{#1}}

\newcommand{\myx}{$\times$\xspace}
\newcommand{\vtt}[1]{\texttt{#1}\xspace}

\newcommand{\fastfair}{FAST \& FAIR\xspace}

\newcommand{\cas}{\texttt{CAS}\xspace}

\newcommand{\store}{\texttt{store}\xspace}
\newcommand{\load}{\texttt{load}\xspace}

\newcommand{\clflush}{\texttt{clflush}\xspace}
\newcommand{\clflushopt}{\texttt{clflushopt}\xspace}
\newcommand{\clwb}{\texttt{clwb}\xspace}

\newcommand{\sysname}{{\textsc{Recipe}}\xspace}

\newcommand{\mfence}{\texttt{mfence}\xspace}

\newcommand{\stateart}{{state-of-the-art}\xspace}

\usepackage{ifthen}
\newboolean{publicversion}
\setboolean{publicversion}{false}
\ifthenelse{\boolean{publicversion}}{
  \newcommand{\grumbler}[3]{}
}{
  \newcommand{\grumbler}[3]{\xspace\textcolor{#3}{\bf #1: #2}}
}

\newcommand{\ra}[1]{\renewcommand{\arraystretch}{#1}}

\setcopyright{acmcopyright} 
\begin{document}


\title{RECIPE : Converting Concurrent DRAM Indexes\\ to
  Persistent-Memory Indexes}

\author{Se Kwon Lee}
\affiliation{%
  \institution{University of Texas at Austin}
}

\author{Jayashree Mohan}
\affiliation{%
  \institution{University of Texas at Austin}
}

\author{Sanidhya Kashyap}
\affiliation{%
  \institution{Georgia Institute of Technology}
}

\author{Taesoo Kim}
\affiliation{%
  \institution{Georgia Institute of Technology}
}

\author{Vijay Chidambaram}
\affiliation{%
  \institution{University of Texas at Austin and VMware Research}
  \vspace{3ex}}

\renewcommand{\shortauthors}{S. K. Lee et al.}

%
%


\copyrightyear{2019}
\acmYear{2019}
\acmConference[SOSP '19]{ACM SIGOPS 27th Symposium on Operating Systems Principles}{October 27--30, 2019}{Huntsville, ON, Canada}
\acmBooktitle{ACM SIGOPS 27th Symposium on Operating Systems Principles (SOSP '19), October 27--30, 2019, Huntsville, ON, Canada}
\acmPrice{15.00}
\acmDOI{10.1145/3341301.3359635}
\acmISBN{978-1-4503-6873-5/19/10}

%

\begin{abstract}
We present \sysname, a principled approach for converting concurrent
DRAM indexes into crash-consistent indexes for persistent memory
(PM). The main insight behind \sysname is that \textit{isolation}
provided by a certain class of concurrent in-memory indexes can be
translated with small changes to \textit{crash-consistency} when the
same index is used in PM. We present a set of \sklee{conditions} that
enable the identification of this class of DRAM indexes, and
\sklee{the actions to be taken to convert each index to be
  persistent}. Based on these \sklee{conditions and conversion
  actions}, we modify five different DRAM indexes based on B+ trees,
tries, radix trees, and hash tables to their crash-consistent PM
counterparts. The effort involved in this conversion is minimal,
requiring 30--200 lines of code. We evaluated the converted PM indexes
on Intel DC Persistent Memory, and found that they outperform
state-of-the-art, hand-crafted PM indexes in multi-threaded workloads
by up-to 5.2\myx. For example, we built P-CLHT, our PM implementation
of the CLHT hash table by modifying only 30 LOC. When running YCSB
workloads, P-CLHT performs up to 2.4\myx better than
Cacheline-Conscious Extendible Hashing (CCEH), the state-of-the-art PM
hash table.
\end{abstract}

\maketitle

\section{Introduction}
\label{sec-intro}

Persistent memory (PM) is an emerging class of memory technology. The
first PM product, Intel DC Persistent Memory, was announced in April
2019~\cite{IntelPMM}. Intel's PM product will be attached to the memory bus and
accessed like DRAM via processor loads and stores. It has a unique
performance profile: read latency 3.7\myx that of DRAM, while read and
write bandwidth are $1/3^{rd}-1/6^{th}$ that of DRAM~\cite{pm-arxiv}.

The low latency and durability of PM make it an attractive medium for
building storage systems. Indexes are key to achieving good read
performance, and are thus a crucial component of several storage
systems. Researchers have designed several PM indexes such
as \fastfair~\cite{hwang2018endurable}, Level
Hashing~\cite{zuo2018write}, CCEH~\cite{nam2019write},
NV-Tree~\cite{yang2015nv}, wB+ tree~\cite{chen2015persistent},
WOART~\cite{lee2017wort}, and FPTree~\cite{oukid2016fptree}. Designing
these indexes from scratch is challenging; the indexes must provide
high performance and concurrency while ensuring that the index
recovers correctly in the event of a power loss or a system
crash. This complexity leads to subtle bugs; for example, we identify
previously unknown data-loss and crash-recovery bugs in the \fastfair
B+ tree and CCEH hash table (\sref{sec-mot}).
\blfootnote{Se Kwon Lee and Jayashree Mohan are supported by SOSP 2019
student travel scholarships from the National Science Foundation and ACM
Special Interest Group in Operating Systems respectively.}

While research on building concurrent, crash-consistent PM indexes has
been gathering traction recently, there have been decades of research
on building concurrent DRAM indexes~\cite{ascy, amp, lock-free-thesis,
ob-free-synch, rscht, lock-free-ll, rcu, lock-free-queues, david2017};
for example, the skip list ~\cite{skiplist89} was invented thirty
years ago. Modern in-memory data structures are carefully designed
keeping in mind cache efficiency~\cite{cache-conscious-db},
pre-fetching~\cite{db-prefetching}, concurrency, and parallelism via
SIMD instructions~\cite{db-simd, indices-survey}. Concurrent data
structures are widely used in industry and academia; for example,
latch-free BwTree in the Hekaton OLTP engine~\cite{hekaton}, Adaptive
Radix Trees in the HyPer database~\cite{hyper}, the Timeline Index in
SAP HANA~\cite{timeline-index}, and Masstree in the Silo database
~\cite{tu2013speedy}.  In this work, we seek to leverage the research
on concurrent DRAM indexes 
to build
persistent PM indexes.

We present \sysname, a principled, practical approach for converting concurrent
DRAM indexes into their persistent, crash-consistent
counterparts. If the source DRAM index that is converted by using the
\sysname approach is correct, the resulting PM index will be correct
as well. We call the index \textit{correct}, if no previously inserted key
is lost and a search returns the latest value of the key.
\sysname can only be applied to DRAM indexes meeting specific
conditions, and the conversion process differs based on the matching
conditions. Therefore,
we introduce a set of \sklee{conditions} that specify which DRAM indexes
can be converted using the \sysname approach; \sklee{DRAM indexes meeting 
these conditions can be converted to their PM counterparts with minimal changes}. We
convert five popular DRAM indexes into their PM counterparts; all
conversions required less than 200 LOC (1--9\% of the codebase). 
Each converted index uses a
different data structure: a hash table, a trie, a B+ tree, a radix
tree, and a combination of tries and B+ trees. The converted PM
indexes offer good performance and scalability, outperforming
hand-crafted \stateart PM indexes on many workloads.

The basic insight behind \sysname is that isolation in concurrent DRAM
indexes is closely related to crash consistency in persistent
indexes. Isolation ensures that reads return correct values and writes
result in consistent states irrespective of other active reads and
writes. We can view crash consistency similarly: reads after a crash
return correct values, and writes after a crash lead to consistent
states (perhaps by fixing inconsistencies). To increase performance,
many DRAM indexes employ non-blocking synchronization. They
allow reads and writes to see inconsistent states; the reads 
and writes have the ability to detect inconsistencies, and
either tolerate or fix them. 
This is the exact set of features required to recover correctly 
after a crash; such DRAM indexes basically have crash-recovery 
logic woven into their reads and writes. We make the observation 
that converting such DRAM indexes into their PM indexes is \sklee{much more} straight-forward \sklee{than designing PM indexes from scratch}; 
since PM and DRAM are both accessed via the same load and store instructions, 
a developer only has to ensure stores are correctly ordered using 
memory fences and flushed from volatile caches to persistent media. 
No new crash recovery algorithms (which tend to be complex) are 
required to be added to the converted PM index, as the reads and 
writes can detect and tolerate or fix inconsistencies already. 

The challenge in developing the \sysname approach is carefully
reasoning about which DRAM indexes can be converted, and how to
convert compatible indexes. For example, read operations in some
lock-free DRAM indexes, on finding an inconsistency \sklee{based on
version numbers}, simply back-off and retry, assuming the
inconsistency is transient. This approach does not work if a crash has
left the index in a permanently inconsistent state. We present
three \sklee{conditions and conversion actions} that precisely capture
the properties the source DRAM index should have, and how to convert
the source DRAM index. The guidance provided by \sysname is not at the
source level, and thus cannot be easily automated. However,
\sklee{we found that the conditions are broadly applicable; we
were able to convert five different DRAM indexes by modifying fewer
than 200 lines of code.}


Building a PM index using the \sysname approach offers several
benefits. First, it drastically lowers the complexity of building a PM
index; the developer simply chooses an appropriate DRAM index and
modifies it as indicated by our approach. The developer does not have
to worry about crash recovery, even in the presence of concurrent
writes. Second, if the developer converts a DRAM index that has high
performance and scalability, the converted PM index also offers good
performance without any further optimization.

\newcolumntype{P}[1]{>{\centering\arraybackslash}p{#1}}

\begin{table}[!t]
  \small
  \centering
  \ra{1.3}
  \begin{tabular}{@{}lp{52pt}P{35pt}ccc@{}}
    \toprule[1.2pt]
    DRAM & Data &  \sklee{\sklee{RECIPE}} &  \multicolumn{3}{c}{Lines of Code} \\
    Index & Structure & Condition & Orig & Core & Modified \\
    \midrule
    CLHT & Hash Table & \#1 & 12.6K & 2.8K  & 30 (1\%) \\
    HOT & Trie & \#1 & 36K & 2K  & 38 (2\%)  \\
    Bw Tree & B+ Tree & \#2 & 13K & 5.2K  & 85 (1.6\%)  \\
    ART & Radix Tree & \#3   & 4.5K & 1.5K & 52 (3.4\%)\\
    Masstree & B+ Tree \& Trie & \#3  & 25K & 2.2K & 200 (9\%) \\
    \bottomrule[1.2pt]
  \end{tabular}
  \vspace{2pt}

\mycaption{Categorizing common DRAM indexes}{The table enumerates popular DRAM indexes, 
the RECIPE \sklee{condition} they satisfy, and the effort required to
convert them to their PM versions. The lines of code in the core
codebase is calculated by excluding tests and helper
libraries.
}
\label{tbl-axiom-categories}
\end{table}

We present our experience with converting five DRAM indexes
using \sysname, each based on a different data structure: Adaptive
Radix Tree (ART)~\cite{leis2016art, leis2013adaptive}, Height
Optimized Trie (HOT)~\cite{binna2018hot},
BwTree~\cite{levandoski2013bw}, Cache-Line Hash Table
(CLHT)~\cite{ascy} and Masstree~\cite{mao2012cache}. 
Table~\ref{tbl-axiom-categories} categorizes these DRAM indexes based 
on the \sklee{condition} they satisfy, and the effort involved in converting 
them to be persistent (as a percentage of change to the core
codebase that excludes tests and helper libraries). 
Masstree was by far the most complicated of these indexes, as it combines tries and 
B+
trees. By applying \sysname, we were able to convert all indexes to
their PM counterparts with 1--9 \% change to the core
source code.

To test the correctness of our converted PM indexes, we introduce a
new methodology for testing crash recovery. We take advantage of the
fact that insert and structure-modification operations (such as a node
split in a B+ tree) in non-blocking indexes are comprised of a small
number of ordered atomic steps. We instrument the code to simulate a
crash in between these atomic steps. Simulating a crash involves
returning from an insert or structure-modification operation mid-way
without cleaning up any state, leaving the index in a partially
modified state. We then continue reading and writing to the index
using multiple threads, testing that the reads return expected values
and writes complete successfully. We also trace all dynamic memory
allocations, stores, and cache line flushes using PIN~\cite{pin}, and
check that all dirtied cache lines in allocated memory ranges are
flushed to PM. Though this method is not exhaustive, it is powerful,
allowing us to find data-loss and crash-recovery bugs in CCEH
and \fastfair. Testing did not reveal any bugs in the PM indexes we
converted.

To test the performance of our converted PM indexes, we use the 
YCSB~\cite{ycsb} benchmark to perform multi-threaded insertions, 
point queries, and range queries on Intel DC Persistent Memory. 
We compare the converted PM indexes against
\stateart manually-designed PM indexes. We find that our
converted PM indexes outperform the \stateart by up-to 5.2\myx in
multi-threaded YCSB workloads. The main performance gain
for \sysname-converted indexes comes from the fact that the DRAM
indexes we convert are already optimized for concurrency and cache
efficiency; the high read latency of PM makes cache efficiency even
more important.

The \sysname approach has a number of limitations. \sysname cannot be
applied to any DRAM index that does not match one of its three
conditions. For instance, \sysname cannot be applied to indexes with
blocking reads or non-blocking reads with version-based
retry. \sysname assumes that the locks used in an index are 
reinitialized to prevent deadlock when an index
recovers from a crash. \sysname assumes garbage collection is employed
for the persistent-memory allocator to reclaim unreachable
objects. \sysname assumes the original DRAM index is correct; if it
has a bug, the converted PM index will also have a bug. 
\errata{Indexes converted by \sysname provide a low level of isolation
(Read Uncommitted); primitives such as the marking-after-flush 
techniques~\cite{wang2018easy, david2018log} and non-temporal stores 
should be used to ensure stronger isolation.} Finally, the main focus 
of this approach is the correct and \sklee{principled} conversion of 
DRAM indexes into PM indexes; there are usually opportunities for 
further optimization. 

In summary, this paper makes the following contributions:
\begin{itemize}
  \item \sysname, a principled approach to convert DRAM indexes into PM indexes 
(\sref{sec-recipe}).
  \item An efficient method for testing crash recovery of
    PM indexes (\sref{sec-test}).
  \item A case study of converting five DRAM indexes based on
    different data structures to PM indexes using the \sysname
    approach (\sref{sec-case}). \sklee{\sysname-converted indexes
    are available at} \url{https://github.com/utsaslab/RECIPE}.
  \item Experimental evidence showing \sysname-converted PM indexes
    recover correctly from crashes, and achieve performance and
    scalability competitive with \stateart hand-crafted PM
    indexes (\sref{sec-eval}).
\end{itemize}

\section{Background}
\label{sec-bkgd}

We begin by describing DRAM indexes and their interfaces, persistent
indexes, and how indexes achieve concurrency and scalability. We then
motivate why a principled approach is required for building PM
indexes.

\subsection{DRAM Indexes}

DRAM indexes are used to efficiently lookup data items in databases,
file systems, and other storage systems. Their interface involves
five main operations:

\vspace{5pt}\noindent\vtt{\bf insert(key, value)} inserts the pair of \vtt{key} and
\vtt{value} into the index. \vtt{value} is usually the location in the
storage system where \vtt{key} can be found.

\vspace{5pt}\noindent\vtt{\bf update(key, value)} update \vtt{key} with
\vtt{value} in the index. Some key-value stores use \vtt{insert} for
both insertions and updates, while other key-value stores will fail
insertions if the key already exists. 

\vspace{5pt}\noindent\vtt{\bf lookup(key)} returns the \vtt{value}
associated with \vtt{key} in the index.

\vspace{5pt}\noindent\vtt{\bf range\_query(key1, key2)} returns all key-value pairs
where the keys are within the specified range. Range queries are
sometimes implemented using an iterator: a cursor that can be
incremented to the next key in the sequence. 

\vspace{5pt}\noindent\vtt{\bf delete(key)} removes the specified key from
the index.

\vheading{Structural Modification Operations (SMOs)}.  SMOs are
operations internal to the data structure, that are required either to
ensure that the invariants of the data structure holds, or to improve
performance. For instance, when the nodes in a B-tree overflow (during
insertion) or underflow (during deletion), node splits or merges are
required to re-establish the invariants of a B-tree. In other data
structures like hash tables, SMOs like re-hashing are necessary to
keep constant average cost per operation.

\vheading{Performance}. DRAM indexes take special care to have high
lookup and insertion performance, as these are often performed in the
critical path. Lookup and insertion performance depend on the number
of processor loads and store required, along with aspects like whether
the layout is cache-friendly and prefetcher-friendly.

\vheading{Correctness}. A DRAM index should return the latest inserted
\vtt{value} for any given \vtt{key}. Unless the key is explicitly
deleted, an inserted key should never be lost. 

\subsection{Concurrency and Isolation}

DRAM indexes use multiple threads to increase throughput on multi-core
machines. However, since all threads operate on the same shared index,
additional mechanisms are required to ensure correctness. Concurrent DRAM
indexes need to provide \emph{isolation}: ensuring that even if
multiple writers are modifying the index at the same time, the final
index state corresponds to the insertions or updates happening in some
sequential order. The index also needs to ensure that reads do not
reflect the result of a partial or incomplete insertion or update
operation.

\vheading{Blocking operations}. The easiest way to ensure correctness
in a concurrent index is to obtain a lock on the index, and only allow threads with lock to
read or write. This serializes all operations and decreases throughput
to that of a single thread. To increase performance, reader-writer
locks are often used~\cite{memcached, rumble2014log, zuo2018write}; 
readers can get a shared lock, all writers have
to contend on a single lock, and there is mutual exclusion between
readers and the writer.

\vheading{Non-blocking operations}. Non-blocking
operations~\cite{tsigas2001evaluating} are employed to fully exploit
the parallelism offered by modern hardware. Non-blocking operations
guarantee progress of some or all remaining threads regardless of the
suspension, termination, or crash failure of one of the
threads~\cite{herlihy1991wait, fraser2007concurrent}.  They provide
consistency and correctness by carefully ordering load and store
instructions using memory fence (\mfence)~\cite{adve1996shared,
  herlihy1990linearizability}, while avoiding the use of mutual
exclusion and expensive synchronization primitives.

Non-blocking operations can be categorized into 
lock-free and wait-free, based on their progress guarantee.
Lock-free operations allow multiple threads to simultaneously 
access a shared object, while guaranteeing that at least one 
of these operations finish after a finite number of
steps~\cite{herlihy1991wait}. Wait-free operations are a subset of
lock-free operations, with the additional condition that every thread
finishes the operation in a finite number of steps~\cite{herlihy1991wait}.

Non-blocking operations are built using hardware-atomic primitives
such as compare and swap (CAS) or test and set. If every update is
performed via a single atomic store, correctness is implicitly
guaranteed. If updates consist of a sequence of atomic stores, then
the readers can either make progress by reasoning about the
deterministic order of stores, or can use additional techniques such
as version-based retry~\cite{fatourou2018efficient, cha2001cache}.

While non-blocking operations are known to provide high performance
and scalability, high contention to the shared resource reduces
performance and could lead to starvation~\cite{faleiro2017latch}.  For
example, if a lock-free write is interrupted by the scheduler, it
might need to retry the operation after being rescheduled if the
shared state has changed.  To protect against starvation, many indexes
use non-blocking reads and blocking, lock-protected
writes~\cite{leis2016art, binna2018hot, mao2012cache, ascy}.

\subsection{Persistent Memory}
Persistent Memory (PM) bridges the gap between DRAM and storage by
offering DRAM-like latency with storage-like persistence.  Writes to
the PM are issued in 8-byte failure-atomic units, which are first
written to the volatile CPU cache. These cache lines can be written
back to the Persistent Memory Controller in an arbitrary order. Intel
x86 architecture provides the \mfence instruction to prevent such
memory reordering~\cite{Intel2019-Arch}; if a store instruction is
followed by a \mfence, then it is guaranteed to be visible before any
other stores that follow the \mfence.  Additionally, to explicitly
flush a cache line to the persistent controller, x86 architecture
provides \clflush, \clwb and \clflushopt instructions.  Our work uses
\clwb and \mfence to guarantee persistence.

\subsection{Crash-Consistent PM Indexes}

Building PM indexes is attractive for two reasons. First, the larger
capacity of PM at close-to-DRAM latencies allows using larger indexes
than possible with just DRAM. Second, DRAM indexes need to be
reconstructed after a crash; for large indexes, reconstruction could
take several minutes or hours. In contrast, a PM index is instantly
available. This has motivated a number of researchers to design
efficient indexes on PM; we count fifteen PM indexes published in top
systems and database conferences since 2015. The PM indexes include
variants of B+ trees~\cite{chen2015persistent, yang2015nv,
  oukid2016fptree, wang2018easy, hwang2018endurable,
  arulraj2018bztree, chi2014making, xia2017hikv, kim2018clfb}, radix
trees~\cite{lee2017wort}, and hash tables~\cite{schwalb2015nvc,
  nam2019write, wu2016nvmcached, zuo2017write, zuo2018write}.

\vheading{Crash Recovery}. One of the main differences between a DRAM
index and a PM index is that the PM index has to ensure that it can
correctly recover in the case of power loss or kernel crash. This
requires carefully ordering stores to PM using \vtt{mfence}
instructions and then flushing the dirty data from volatile caches to
persistent media using cache line flush instructions (\clflush, \clwb,
or \clflushopt )~\cite{pelley2014memory}. If the write is larger than
eight bytes, a crash could lead to a torn write where the data is
partially updated; techniques such as logging~\cite{hagmann87-cedar}
and copy-on-write~\cite{wafl} are used to provide atomicity.

\section{Motivation}
\label{sec-mot}

Hand-crafted PM indexes employ non-blocking operations to increase
scalability~\cite{hwang2018endurable, nam2019write}.  However, while
non-blocking operations offer high performance and scalability, their
complexity makes it challenging to develop, test, and debug indexes
with non-blocking operations.  Persistence makes the problem even
harder, since developers have to ensure that crash recovery and
concurrency mechanisms interact correctly. We analyze two
state-of-the-art PM indexes: the \fastfair B+
tree~\cite{hwang2018endurable}, and the CCEH hash
table~\cite{nam2019write}.

\vheading{\fastfair}. \fastfair is a PM B+ tree that provides
lock-free reads. The reads detect and tolerate inconsistencies such as
duplicated elements in a sorted list. Writers hold a lock for mutual
exclusion. The writes detect inconsistencies such as duplicated
elements, and try to fix them. However, we found that concurrent
writes could lead to loss of a successfully written key.

Consider the following scenario. Two threads try to insert keys to the
same internal node concurrently; one thread gets a lock and performs a
node split. When the other thread gets the lock, it does not realize
the node has been changed, and inserts the key into the wrong
node. The insert is successful, but a reader would never be able to
find the inserted value. We confirmed this design-level bug with the
\fastfair authors. The solution is to add metadata about the high-key
to B+ tree nodes, as done by prior works~\cite{cha2001cache,
  levandoski2013bw, mao2012cache}. Please refer to our bug report for
more details~\cite{fair-bug-1}.

We also found an implementation bug. According to its design, \fastfair
 recovers correctly from crashes at any point, not losing any
inserted keys. However, when we crashed \fastfair consecutively 
in the middle of split and merge operations on two nodes, 
keys present in the right node were
lost. This is a testament to the complexity of these indexes; a
correct design is not always translated properly to a correct
implementation. 

Finally, we found that incorrect crash recovery can result in poor
performance. If \fastfair crashes in the middle of splits, although
the recovered structure is correct, it is not efficient. A series of
such crashes transforms the B+ tree into a linked list, leading to
poor read and write performance.

In summary, our investigation of \fastfair revealed a design-level
bug that lost data, an implementation-level bug that lost data, and
that crashes can lead to poor performance. The design-level bug
resulted from not leveraging prior research on concurrency, where the
high-key problem and its solution is well-known.

\vheading{CCEH}. We discovered that the CCEH PM hash
table~\cite{nam2019write} has two bugs: one in its directory doubling
code, and one in crash recovery code. Directory doubling is similar to
rehashing the hash table. There are three pieces of metadata that CCEH
has to atomically update in correct order during directory doubling:
the pointer to the directory, the directory width, and the global
depth. If a crash happens before the global depth is updated,
insertion operations loop infinitely. If a crash happens after the
pointer to the directory is swapped, the crash recovery algorithm goes
into an infinite loop. The authors of CCEH have acknowledged both
bugs.

\vheading{Summary}. We find the ad-hoc design of concurrent,
crash-consistent PM indexes makes it hard to reason about behavior
during concurrent writes and crashes, leading to bugs. There is a need
both for principled design of PM indexes, and testing whether PM
indexes correctly recover from crashes.



\section{The \sysname Approach}
\label{sec-recipe}

We present \sysname, a principled approach for converting a specific
class of DRAM indexes to their crash-consistent PM counterparts. The
converted PM index inherits correctness and scalability from the DRAM
index. The \sysname approach guarantees that the converted PM index
will recover from crashes correctly. Thus, if the developer uses the
\sysname approach to convert an appropriate DRAM index, the resulting
PM index will be correct, concurrent, and crash-consistent.

\sysname \sklee{identifies three categories of DRAM indexes} to guide
this conversion.  Each \sklee{category is accompanied by a condition
  and conversion action} of the form: ``\emph{if the DRAM index
  satisfies these conditions, then convert it to a PM index using
  these conversion actions}''. We first present the intuition behind
the \sysname approach, and then describe each \sklee{category}.

\subsection{Overall Intuition}


We observe that some DRAM indexes use non-blocking reads (such as
lock-free reads) to improve performance. These non-blocking reads may
observe inconsistent states since writes may be underway at the time
of read; the read operations can then \emph{tolerate} such
inconsistencies, returning a consistent answer to the user.  For
example, the read operation may see duplicate records and only return
a single record to user~\cite{hwang2018endurable, gfs}. Similarly,
write operations may also see an inconsistent state and \emph{fix} the
inconsistency; write operations in BwTree perform such
fixes~\cite{levandoski2013bw}. Prior theoretical work has termed this
a \emph{helping mechanism}, where an operation started by one thread
which fails is later completed by another thread~\cite{help}.
  
The \sysname approach is based on the following insight: if reads can
tolerate inconsistencies, and writes can fix them, a separate
crash-recovery algorithm is not required. DRAM data structures that
have such read and write operations are \emph{inherently
  crash-consistent}. If such data structures are stored on PM instead
of DRAM, they would be crash-consistent with minimal modifications;
the developer would only need to ensure that all data dirtied by store
operations are persisted to PM in the right order. We refine this
observation through three \sklee{conditions with corresponding
  conversion actions} that help a developer convert a DRAM index into
a crash-consistent, concurrent PM index.

\subsection{Assumptions and Limitations}

\sysname assumes that the locks used in the index are non-persistent,
and that the locks are re-initialized after a crash (to prevent
deadlock). \sysname also assumes that unreachable PM objects will be
garbage collected, as a failed update operation may result in an
allocated but unreachable object. Finally, \sysname also assumes that
the DRAM index operates correctly in the face of concurrent writes.

\sysname can only be applied to DRAM indexes that match one of the
three conditions. For example, DRAM indexes which employ blocking
reads or non-blocking reads with retry mechanisms cannot be converted
using \sysname. The three conditions which follow specify precisely
which DRAM indexes can be converted by \sysname.

\subsection{Condition \#1: Updates via single atomic store}
\label{axiom1}

Reads must be non-blocking, while writes may be blocking or
non-blocking. The index makes write operations visible to
other threads using a single hardware-atomic store.

\vheading{Conversion Action}. Insert cache line flush and memory fence 
instructions after each \vtt{store}. \errata{For non-blocking writes 
where the mismatch between the store orders to CPU cache and PM can occur,
each load should also be followed by cache line flush and memory fence instructions.}

\begin{figure}[!t]
  \centering
  \includegraphics[width=0.4\textwidth]{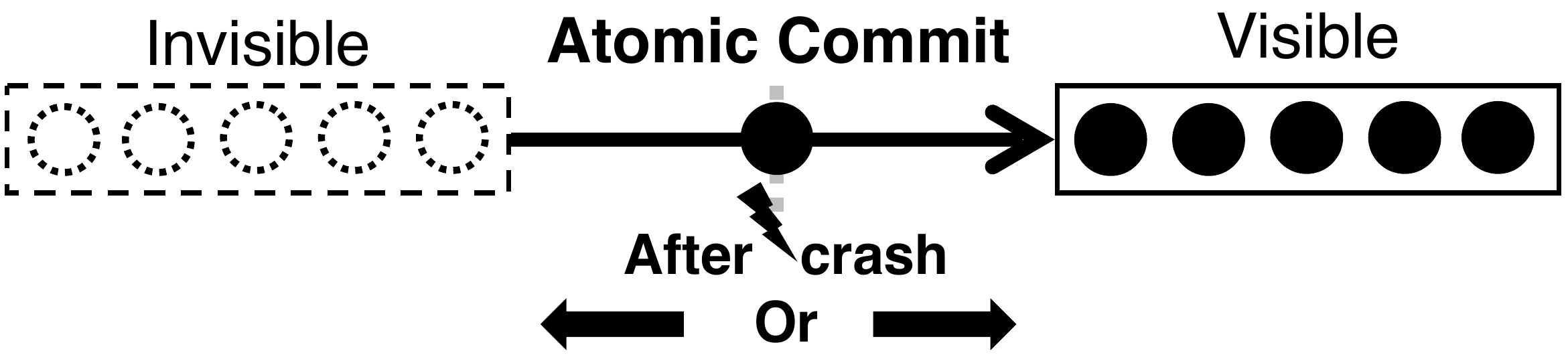}
  \vspace{-5pt}
  \mycaption{\sklee{Condition 1}}{When a crash occurs in the middle of
    an update operation that completes in a single hardware atomic
    step, there is no recovery required. The state after the crash is
    either the initial state or the final state.}
  \label{fig-axiom1}
  \vspace{0pt}
\end{figure}

Fig~\ref{fig-axiom1} illustrates the scenario covered by \sklee{Condition} \#1. 
The index moves in a single atomic step from its initial state to
its final state.
A crash at any point leaves the index consistent, so crash recovery is not required.

\errata{If writes are blocking,} converting an index that fits these conditions 
is straight-forward; each store instruction must be followed by a cache line 
flush and a memory fence instruction. This ensures that all dirty data is flushed
to PM, and that the order in which the writes happen in CPU cache is
the same order in which they are persisted to PM \errata{within critical sections
protected by locks.}

\errata{However, if writes are non-blocking, the stores to PM can be reordered 
because the cache line flushes from concurrent writers cannot be synchronized without locks. 
If the next writer performs new updates dependent on the prior write operation not yet persisted,
the new updates can be lost after a crash. Therefore, each load and store should also be 
followed by a cache line flush and a memory fence~\cite{izraelevitz2016linearizability} 
to enforce the stores to CPU cache and PM to be the same order.}

Performance can be increased by allowing stores preceding the final critical
store to be reordered~\cite{pelley2014memory}. Instead of putting a
fence after each store, we would need fences only surrounding the
final atomic store.

\vheading{Examples}. We converted two indexes, the Height Optimized Trie (HOT), 
and the Cache-Line Hash Table (CLHT) based on \sklee{Condition} \#1.
These indexes employ copy-on-write for updates and failure-atomically make
them visible to other threads via atomic pointer swap while protected by
locks. Thus, their conversions just require adding cache line 
flushes and memory fences after each store.

\subsection{\sklee{Condition \#2}: Writers fix inconsistencies}
\label{axiom2}

Reads and writes must be both non-blocking. The index performs write
operations using a sequence of ordered hardware-atomic stores. If the
reads observe an inconsistent state, they detect and \emph{tolerate}
the inconsistency without retrying. If writes detect an inconsistency,
they have a \emph{helping mechanism} which allows them to fix the
inconsistency.

\vheading{Conversion Action}. Insert cache line flush 
and memory fence instructions after each \vtt{store} \errata{as well as
\vtt{load} to ensure that a previous state is persisted first}.
\begin{figure}[!t]
  \centering
  \includegraphics[width=0.45\textwidth]{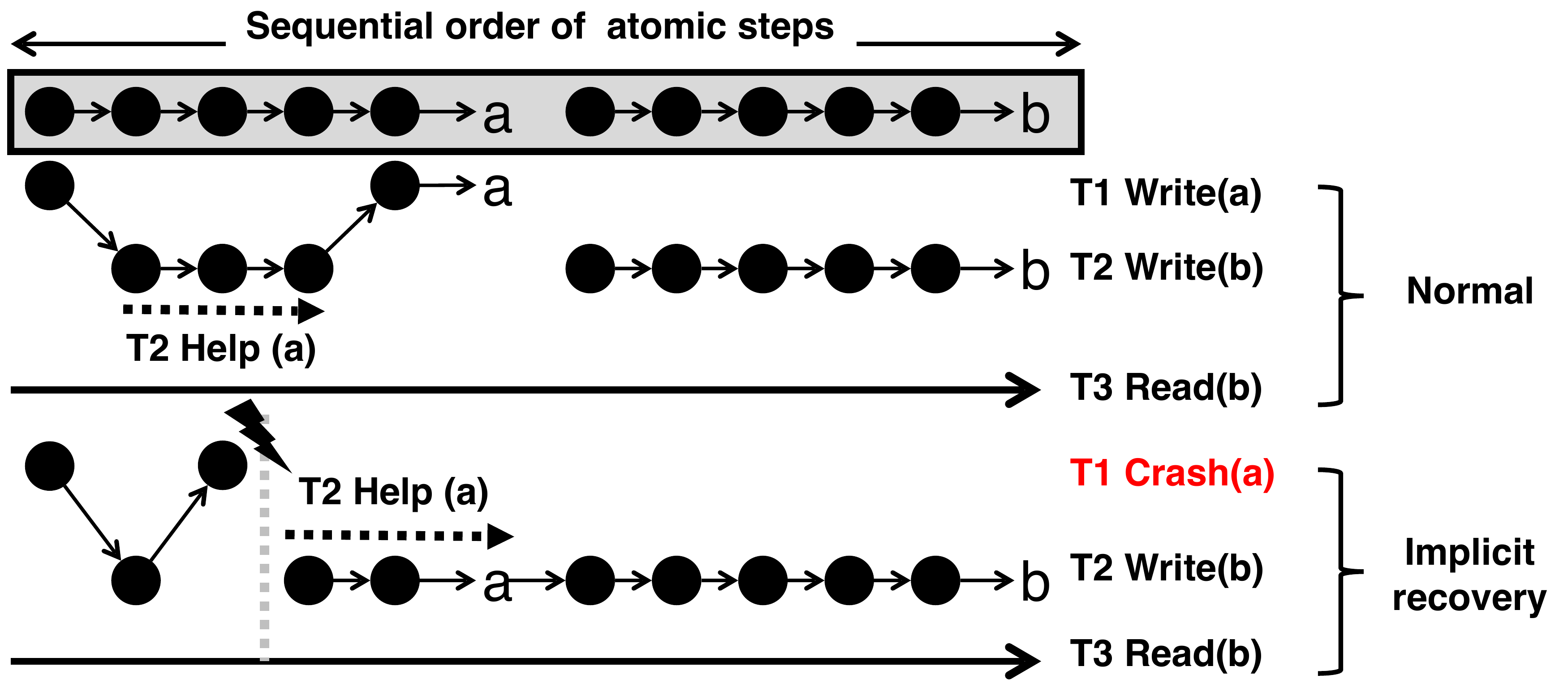}
  \mycaption{\sklee{Condition 2}}{A crash occurs in the middle of
    Thread 1's write operation. Thread 2 detects this, completes
    Thread 1's write operation using its helper mechanism, and then
    proceeds with its own write operation.}
  \label{fig-axiom2}
  \vspace{0pt}
\end{figure}

Figure~\ref{fig-axiom2} illustrates the scenario for \sklee{Condition
\#2}. A crash leaves Thread 1's write partially completed. Thread 2 is
able to detect this; since the write operation comprises of a small
sequence of deterministic steps, Thread 2 can identify where the crash
happened. Thread 2 then proceeds to complete the operation, and then
proceed with its own write. This restores the index back to its
consistent state. Any read observing these actions is able to tolerate
the inconsistency and return a consistent value back to the user.

Note that in general, it is hard after a crash to identify what
happened before the crash if extra information is not logged. Indexes
meeting Condition \#2 are able to do this because write operations in
such indexes are comprised of a small number (typically fewer than
five) of ordered store operations which mutate the index in a
deterministic fashion. Thus, after a crash, the write operation can
always deduce what happened before a crash.

Indexes matching \sklee{Condition \#2} do not need any explicit
crash-recovery code because \emph{implicit crash recovery} is already
part of the read and write operations. The first writer that tries to
update the index after a crash and detects the inconsistency is
responsible for the recovery of the part of the index the writer deals
with.  As a result, every store instruction should be followed by a
cache line flush and a memory fence instruction. \errata{However,
the stores to PM can be reordered like the non-blocking writes in Condition \#1 
while concurrent threads cooperate in helping mechanism~\cite{izraelevitz2016linearizability, wang2018easy}.
Therefore, the memory address referred to by the load instructions in helping mechanism
should also be flushed to ensure that a prior state is persisted first.}

\vheading{Example}. The BwTree has non-blocking read and write
operations. It uses a sequence of ordered 
atomic stores to perform Structural Modification
Operations (SMO) like node splits and merges. BwTree write operations
have helper mechanisms which complete and commit any intermediate SMO
state encountered, before proceeding with their own write. Thus, BwTree 
fits into \sklee{Condition \#2}, and we converted it to its persistent version
simply by adding cache line flushes and memory fences.

\subsection{\sklee{Condition \#3}: Writers don't fix inconsistencies}
\label{axiom3}

Reads must be non-blocking, while writes must be blocking.  Write operations
involve a sequence of the ordered atomic steps similar to Condition \#2,
but they are protected by write exclusion (locks).  Reads can detect
and tolerate inconsistencies. Writes can detect inconsistencies;
however, they lack the helper mechanisms needed to fix the
inconsistency.

\begin{figure}[!t]
  \centering \includegraphics[width=0.45\textwidth]{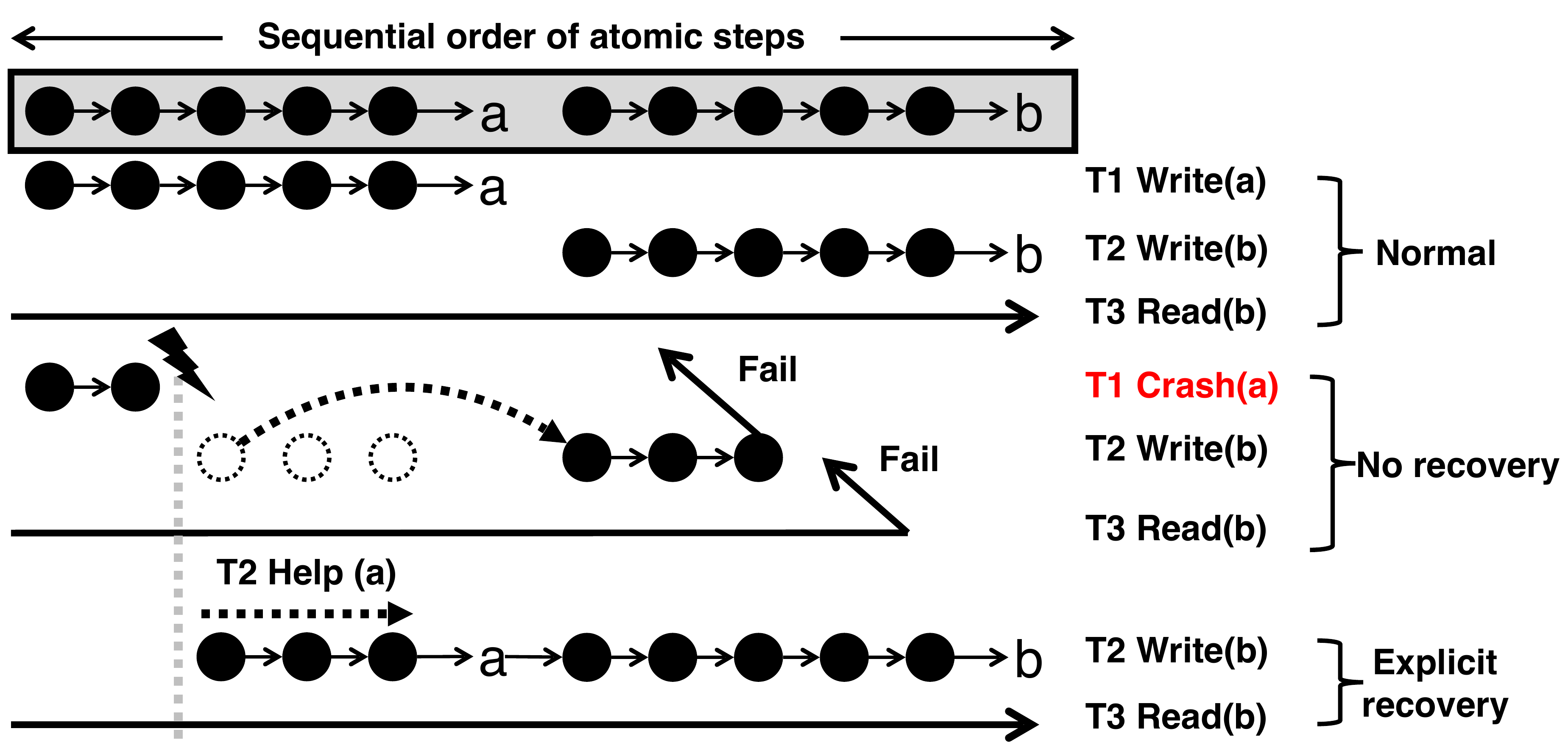}
  \mycaption{\sklee{Condition \#3}}{Condition \#3 indexes lack the
    helper mechanism which allows them to resume an interrupted write
    operation. We explicitly add the helper mechanism which identifies
    that Thread 1's write operation was interrupted, and finishes the
    write operation before proceeding with Thread 2's write operation.
  }
  \label{fig-axiom3}
  \vspace{0pt}
\end{figure}

\vheading{Conversion Action}. Add mechanism to allow writes to detect
permanent inconsistencies. Add helper mechanism to allow writes to fix
inconsistencies. Insert cache line flush 
and memory fence instructions after each \vtt{store}.

Indexes conforming to \sklee{Condition \#3} are the hardest to
convert, as they require multiple steps. The root of the problem is
that \sklee{Condition \#3} indexes do not have helper mechanisms in
their write operations. Therefore while reads and writes tolerate
inconsistencies, the permanent inconsistency will never get fixed.

First, the write operation must distinguish between a transient
inconsistency due to another on-going write or a permanent
inconsistency due to a crash. It differentiates these scenarios by
trying to acquire the write lock; if it is successful, there are no
other writes happening concurrently, so an inconsistent state must be
due to a crash.

Second, a helper mechanism must be added to finish an interrupted
write operation. We find that helper mechanism can be built using code
from the write path. The helper mechanism must first identify what was
happening at the point of the crash (similar to Condition \#2); it
must then complete the interrupted write
operation. Figure~\ref{fig-axiom3} illustrates that explicit recovery
code must be added into the writer for \sklee{Condition \#3} indexes.

Adding the helper mechanism to write operation is correct since it
re-uses code from the write path; reads can already tolerate the
inconsistencies due to on-going writes. Adding the helper mechanism
converts a Condition \#3 index into a Condition \#2 index. At this
point, only adding cache line flushes and memory fences after each
store are required to produce a crash-consistent, concurrent PM index.

\vheading{Example}. The Adaptive Radix Tree (ART) falls into the
category of \sklee{Condition \#3}. The writes in ART do not have the
helper mechanism, \sklee{so they just tolerate inconsistencies, when
  encountering an intermediate state of Structural Modification
  Operations (SMO)}. Fortunately, ART's SMO consist of exactly two
ordered steps; after a crash, the helper mechanism only needs to
identify if step one or two has occurred. We modified ART to introduce
\sklee{permanent} inconsistency detection and helper mechanisms, along
with adding cache line flushes and memory fences.

\section{Testing Crash Recovery of PM Indexes}
\label{sec-test}

We introduce a novel method to test whether PM indexes recover
correctly after crashes. Testing crash recovery involves testing two
things: whether the PM index recovers to a consistent state, and
whether the PM index loses any data successfully persisted before the
crash. Consistency for a PM index involves reads and range queries of
all previously inserted keys returning the correct values, and further
writes completing successfully.

The main challenge in testing crash recovery is deciding where to
crash in each workload. A crash could happen after each 8-byte
atomic store in a workload; this makes the total space of crashes in a
reasonable workload prohibitively large. We address this challenge by
observing that most operations in PM indexes are comprised of a small
number of atomic stores; it is enough to simulate a crash after
each atomic store. For each operation in a PM index, we simulate a
crash after all its atomic stores. This is feasible since PM
indexes have few operations, and each operation has few atomic
steps. Structure modifications operations and insertions have less
than five atomic steps in all the PM indexes we tested. Thus, crashing
only after atomic stores drastically reduces the search
space. While there are existing tools like PM-Inspector~\cite{pminspector},
\vtt{pmreorder}~\cite{pmreorder}, and \vtt{yat}~\cite{yat} to simulate crashes, these tools still
pick crash points in a random or exhaustive manner; our targeted
crashing strategy is powerful, revealing bugs with limited testing.

\vheading{Testing consistency}. We test for consistency using three
steps. First, we run a write-heavy workload, and probabilistically
simulate a crash after an atomic store in either insertion or a
structure modification operation like a node split. A crash is
simulated by returning from the operation without any
clean-up activities, leaving a partially modified state. Next, we
explicitly call the recovery function if the PM index has one. We
perform a number of read and write operations using multiple threads,
keeping track of all successfully inserted keys. Finally, we read back
all successfully inserted keys and check that they have the right
values. Note that this approach does not require actual PM; we are
able to emulate crashes using DRAM.

\vheading{Testing durability}. Testing durability involves checking
that all cache lines which were dirtied during the workload are
flushed to PM. This ensures that data written to the PM index is not
lost if there is a crash. To test durability, we use the
Pin~\cite{pin} tool to trace all allocations made using \vtt{malloc},
\vtt{posix\_memalign}, and \vtt{new}. We then trace all store
instructions to these allocated regions, and verify that all dirtied
cache lines are safely flushed to PM. We perform this testing using
two phases: a load phase and a test phase. We first load the index
with enough keys such that future insertions will trigger node splits
and other structure modification operations. In the test phase, we
perform the insertion while tracing allocation, stores, and
cache line flushes. For each insertion, we verify that all dirtied cache lines
were persisted. 

\section{Case Studies}
\label{sec-case}
\begin{table}[!t]
  \small
  \centering
  \ra{1.3}
  \begin{tabular}{@{}llcccc@{}}
    \toprule[1.2pt]
    DRAM & \multicolumn{2}{c}{Synchronization} & \multicolumn{2}{c}{\sklee{Conditions}} \\
    Index & Reader & Writer & Non-SMO & SMO \\
    \midrule
    CLHT & Non-blocking & Blocking & \#1 & \#1 \\
    HOT & Non-blocking & Blocking & \#1 & \#1 \\
    BwTree & Non-blocking & Non-blocking & \#1 & \#2 \\
    ART & Non-blocking & Blocking & \#1 & \#3 \\
    Masstree & Non-blocking & Blocking & \#1 & \#3 \\
    \bottomrule[1.2pt]
  \end{tabular}
  \vspace{2pt} \mycaption{Categorizing convertion actions}{The table
    lists the converted DRAM indexes with their category and
    synchronization properties.}
  \label{tbl-conversion-actions}
  \vspace{-10pt}
\end{table}

We describe how we modified five concurrent DRAM indexes to their PM
counterparts. For each index, we discuss and modify the main write
operations of the indexes in accordance with the proposed conversion
actions.  The operations we modify are classified into Structural
Modification Operations (SMOs) and Non-SMOs (Inserts and Deletes).
Non-SMOs affect a single node (in tree based indexes) or a single
bucket (in hash tables), whereas SMOs require changes to multiple
nodes or buckets.  Table~\ref{tbl-conversion-actions} lists the
converted indexes along with their categories and properties.

\vheading{Lock initialization}. Some of the converted indexes use
locks for write exclusion. These locks are embedded into the node or
bucket structure and are persisted along with the node. However, locks
are required only to provide concurrency; persisting them can result
in deadlocks if a system crash occurs. We re-initialize locks on
startup for all indexes converted using \sysname. We statically
allocate a lock table that holds pointers to each node's lock. This
lock table is initialized when the PM index is restarted after crash.

\vheading{Crash detection}. When a converted PM index detects an
inconsistency during path traversal, it tries to acquire the lock for
the node using \vtt{try lock}. If it fails to acquire the lock, either
the inconsistency is transient due to a concurrent write, or another
write operation is in the process of fixing the inconsistency. If the
write operation acquires the lock, it fixes the inconsistency using
the helper mechanism.

\subsection{Trie: Height Optimized Trie (HOT)}

The Height Optimized Trie (HOT) is a lookup and space-optimized
variant of a trie, where the children of each node in the search tree
share a prefix of the key.  HOT achieves cache efficiency, dynamically
varying the number of prefix bits mapped by a node to maintain
consistent high fanout. The layout is
designed for compactness and fast lookup using SIMD instructions.

\vheading{Non-SMOs}.  HOT uses copy-on-write and commits an insert or
delete operation by atomically swapping the single parent pointer per
operation.  It uses non-blocking read and exclusive write to prevent
the updates from getting lost due to competing pointer swap
operations.

\vheading{SMOs}.
SMOs in HOT occur when prefix bits are mismatched.
If SMOs are required during insertion and
deletion, HOT first identifies the set of nodes to be modified, locks
them bottom up to avoid deadlock, performs the update using
copy-on-write and then unlocks them top down.

\vheading{Conversion to PM.} HOT abides by \sklee{Condition \#1}
because every update to the index is installed through an atomic
pointer swap. Therefore, as long as the \store instructions are
correctly ordered and flushed, crashes will not result in
inconsistencies. Conversion to P-HOT required adding 38 LOC ($<$2\% of
the 2K LOC in HOT core).

\subsection{Hash Table: Cache-Line Hash Table (CLHT)}

CLHT is a cache-friendly hash table that restricts each bucket to be
of the size of a cache line (64 bytes). 
At most three key-value pairs, whose keys 
and values are 8 byte each, fits into one bucket. The design aims at
addressing the cache-coherence problem by ensuring that each update to
the hash table requires one cache line access in the common case. To
ensure that a non-blocking reader finds the 
correct value, CLHT uses atomic snapshots of key-value pairs~\cite{ascy, david2014designing}

\vheading{Non-SMOs}. 
CLHT installs any update to the hashtable by locking the appropriate bucket, 
performing the update in-place and then unlocking it. 
CLHT installs the insert and delete operation 
using a single atomic commit point, ordered by memory fences: 
writing the correct value first prior to updating 8 byte key 
(for insertion) and writing 0 to the key (for deletion).

\vheading{SMOs}. If the inserts extend the number of buckets per
hash beyond a threshold, CLHT performs re-hashing using copy-on-write.
The old hash table is first locked for write. The entries in 
each bucket are then copied over to the new hash table, and finally, 
the old hash table is atomically swapped with the new one.

\vheading{Conversion to PM.} CLHT abides by \sklee{Condition} \#1
because the inserts, deletes, and re-hashing are effected via a single
atomic store. Similar to HOT, we insert cache line flushes and memory
fences after appropriate \store instructions to build
P-CLHT. Common-case non-SMOs (inserts and deletes), except for
re-hashing, require only one cache line flush per update. Conversion
involved 30 LOC (CLHT lock-based implementation is 2.8K LOC).

\subsection{B+ TREE: BwTree}

BwTree is a variant of B+ tree that provides non-blocking reads and
writes.  It increases concurrency by prepending delta records
(describing the update) to nodes. It uses a mapping table that enables
atomically installing delta updates using a single Compare-And-Swap
(\cas) operation. Subsequent reads or writes to this node replay these
delta records to obtain the current state of the node.

\vheading{Non-SMOs}.  Insert and delete operations prepend the delta
record to the appropriate node, and update the mapping table using
\cas. If a \cas to the mapping table fails because of another
concurrent update, the thread simply aborts its operation and restarts
from the root.

\vheading{SMOs}. 
When the base node in BwTree overflows (or underflows), a node split (or merge) is necessary.
BwTree uses a helper mechanism~\cite{levandoski2013bw} to co-operatively
perform concurrent updates in the presence of structural modifications
due to node splits and merges. Any subsequent writer thread that
observes an ongoing split or merge operation first tries to complete
it, before going forward with its own operation.

Splits and merges first post a special delta record to the node to
indicate that a modification is in progress. It then uses the two-step
atomic split mechanism of B-link trees~\cite{lehman1981efficient} to
create a new sibling node in the first step and later update the split
key in the parent node. For node merges, the left sibling of the node
to be merged is updated with a physical pointer to this node and then
the merge key in the parent is removed.

\vheading{Conversion to PM}. \errata{BwTree's non-SMOs are completed by
preprending a new delta node and making it visible with 
a single \cas on the global indirect pointer in the mapping table.
Therefore, BwTree's non-SMOs fit into \sklee{Condition} \#1. 
However, \load operations in non-SMOs of BwTree do not need 
to be followed by cache line flushes, since the stores to PM of non-SMOs 
eventually reflect the same store orders to CPU cache. It is because all 
non-SMOs to the same node by multiple concurrent writers are invisible 
until a single \cas on the same indirect pointer. Even if cache line 
flushes to the same indirect pointer are reordered, the first cache line 
flush persists the most up-to-date \cas correctly. Therefore, for non-SMOs 
of P-BwTree, we perform a cache line flush only if the \cas succeeds while 
not flushing after each \load operation.}

BwTree's node split and merge mechanisms expose intermediate
states to other readers and writers. While readers never restart in
the original design of the BwTree, the open-source implementation of
BwTree allows reads to restart if a node merge is in
progress~\cite{wang2018building}.  We address this issue by modifying
the reader to avoid retry using the inconsistency detection and fix
algorithm already present in the write path of BwTree.

Using their helper mechanism, the writers in BwTree detect and fix any
partially completed operation. As a result, SMOs of BwTree (after
modifications to the read operation) fits into \sklee{Condition} \#2. We
build SMOs of P-BwTree by adding cache line flushes and memory fences
after every \store and \errata{\load operation} to the nodes and mapping table. 
Building P-BwTree involves modifying 85 LOC, as compared to 5.2K LOC in the
core BwTree index.


\subsection{Radix Tree: Adaptive Radix Tree (ART)}
ART is a radix tree variant that reduces space consumption by
adaptively varying node sizes based on the valid key entries. 
8-bit prefix (one byte) is indexed by each node.
The 8-byte header of each node in ART compresses some part 
of common prefix and the length of it. The level field
in each node represents the full length of common prefix 
shared at this node and is never modified after its creation. 
As in HOT, synchronization is provided using non-blocking read and 
exclusive write~\cite{leis2013adaptive, leis2016art}.

\vheading{Non-SMOs}.  For an insertion, a new key-value pair is
appended into the end of the entries in a node and is atomically made
visible by increasing counter value. Deletion is completed via a
single atomic store, simply invalidating a key by setting the value
entry to be NULL.  If the node overflows (or underflows), the node is
copied to a new larger (or smaller) node and then the parent pointer
is atomically swapped.

\vheading{SMOs}. If 
two keys share the same prefix, ART compresses the native radix 
tree structure by simply storing the common prefix in a single 
node (instead of allocating a node per character in the key). 
As key distribution varies, the compressed prefix could 
be expanded or compressed, resulting in split or merge of
existing nodes. Unlike non-SMOs, 
these structural changes are installed in multiple atomic steps. 
If the insertion of a key requires a path compression split, a new
node pointing to the key is first installed, and then the header is
updated to contain the correct prefix.

\vheading{Conversion to PM.}  Since non-SMOs are always committed
atomically, they abide by \sklee{Condition} \#1. However, the path
compression mechanism in ART exposes intermediate states which reads
can tolerate. A read counts the depth of the native decompressed radix
tree while traversing tree, and compares level field with the sum of
the depth and the prefix length stored in a node; if there is a
mismatch, the read simply ignores a part of the prefix at this node to
access the correct key. To ensure correctness, reads verify if the
retrieved key is same as the search key before returning. Writes
similarly detect inconsistencies, but do not fix them.

To build P-ART, we modify the write path to include crash detection
and recovery.
When the node traversal in the write path detects an inconsistency, it
first checks for a crash using a \vtt{try lock}. If it successfully
acquires the lock, the write calculates and persists the correct
prefix. Implementing these changes, along with insertion of cache line
flushes and memory fences required adding 52 LOC to the 1.5K LOC of
ART.


\subsection{Hybrid Index: Masstree}
\label{sec-case-masstree}
Masstree is a cache-efficient, highly concurrent trie-like concatenation of B+ tree
nodes~\cite{mao2012cache}. Masstree provides synchronization 
using write exclusion and lock-free readers retry
when inconsistencies are detected by using version numbers. 

\vheading{Non-SMOs}.  Similar to ART, the non-SMOs of Masstree start
with non-blocking tree traversal to return correct leaf node.  Inserts
to the leaf nodes in Masstree are performed by appending a new
key-value pair to the node with unsorted order and atomically
switching to an updated copy of the 8-byte permutation table,
specifying the sorted orders of keys and empty entries. For deletes,
it is sufficient to atomically update the permutation table to
invalidate the entry.

\vheading{SMOs}.  The internal nodes in Masstree maintain keys in
sorted order using a non-atomic key-shifting algorithm which exposes
inconsistent data to readers~\cite{chen2015persistent}. Reads
therefore retry until the ongoing operation completes. Node splits and
merges lock the corresponding nodes and update version counters upon
completion. Meanwhile, all concurrent reads and writes to these nodes
would simply retry from the root.

\vheading{Conversion to PM.}  The non-SMOs of Masstree abide by
\sklee{Condition} \#1, since insertions and deletions are atomically
reflected by updating a permutation table.  However, Masstree SMOs do
not directly fit into our \sklee{conditions} as the readers do not
tolerate inconsistency without restarts. While the structure of leaf
nodes allows a 2-step atomic split mechanism, the internal nodes do
not. Therefore, we modify the internal nodes to resemble the leaf
nodes, modifying the data structure to resemble the B-link Tree.  This
modification allows a 2-step atomic split mechanism across all levels.
For example, if the insertion requires node split, half of the entries
in split node are copied into the new sibling node, and then the
sibling pointer of split node is atomically installed to the new
sibling. Finally, the entries copied into new sibling node are
atomically invalidated from split node by updating 8-byte permutation
table. Furthermore, this eliminates restarts at the read path.
All the intermediate states exposed by SMOs are tolerated by moving
towards next sibling node, utilizing the B-link Tree's sibling link
and high key~\cite{cha2001cache}.  Reads therefore always return
consistent data and writes can reach the correct leaf node without
retry.

With this change, SMOs of Masstree fits into \sklee{Condition} \#3,
where reads return consistent values, but writers have no mechanism to
fix inconsistent states.
We implement write path recovery by simply replaying the node split
algorithm whenever a crash is detected using a \vtt{try lock}.  If the
intermediate state observed was due to a node split, then this action
would complete the split operation. If the observed crash state was
due to a node merge, replaying the split operation will undo the
merge, bringing the index back to a consistent state.


\section{Evaluation}
\label{sec-eval}

\begin{table}[!t]
  \small
  \centering
  \ra{1.3}
  \begin{tabular}{@{}lll@{}}
    \toprule[1.2pt]
    Workload  & Description  &  Application pattern \\
    \midrule
    Load A & 100\% writes & Bulk database insert\\
    A &  Read/Write, 50/50 & A session store\\
    B & Read/Write, 95/5 &  Photo tagging\\
    C & 100\% reads & User profile cache\\
    E & Scan/Write, 95/5 &  Threaded conversations\\
    \bottomrule[1.2pt]
  \end{tabular}
\vspace{1pt} \mycaption{YCSB workload patterns}{The table describes
  different workload patterns from the YCSB test suite.}
  \label{tbl-ycsb}
\end{table}

\begin{figure*}[!t]
  \centering 
    
  \subfloat[Integer keys : Multi threaded YCSB]{{\includegraphics[width=.475\linewidth]{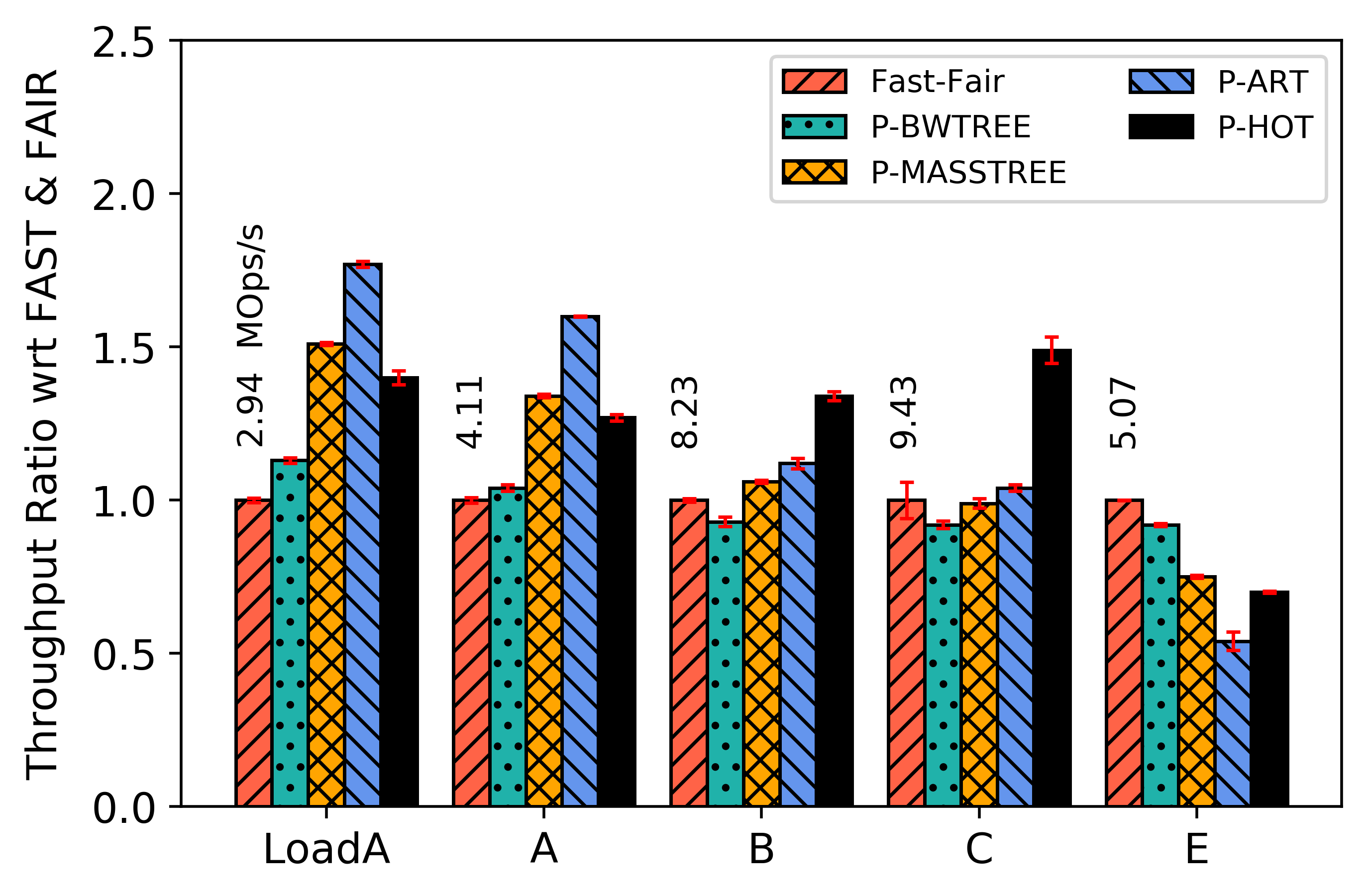} }}%
  \qquad
  \subfloat[String keys : Multi threaded YCSB]{{\includegraphics[width=.475\linewidth]{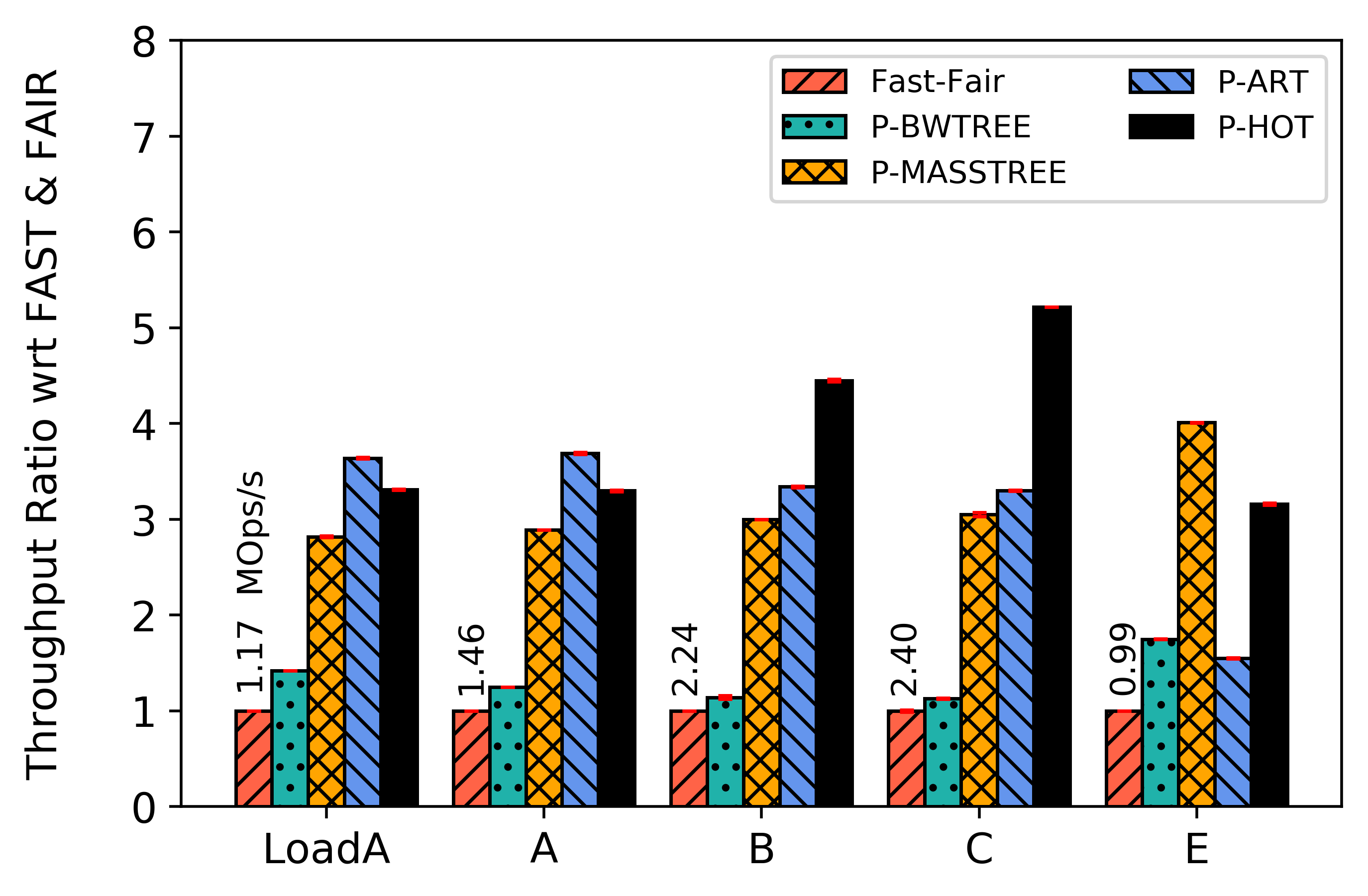} }}%
  
  \subfloat[Integer keys : Performance counters]{%
	\centering
 	 \begin{tabular}[b]{@{}l|cc|ccccc@{}}
    	\toprule[1.2pt]
    	 \multirow{2}{*}{PM Index} & \multicolumn{2}{c|}{Instructions} & \multicolumn{5}{c}{Last Level Cache Miss}\\
    	\cline{2-8}
	  &\clwb & \mfence & LoadA & A & B & C & E \\
    	\midrule
   	 \fastfair &   7 & 8 & 11 & 10 & 8 &  7  & 8 \\
    	P-Bw Tree &  7 & 4 & 17 &  15 &  10 & 9 & 26\\
    	P-Masstree & 3 &  5 & 7 & 7 & 6 & 5  & 8 \\
    	P-ART &  3  &  3 &  4 & 4 & 4  & 4 & 12 \\
    	P-HOT & 7 & 5 & 4 & 4  & 2 & 2  & 10 \\
    	\bottomrule[1.2pt]
 	 \end{tabular}
}
\qquad
\subfloat[String keys : Performance counters]{	 
	  \centering
 	  \begin{tabular}[b]{@{}l|cc|ccccc@{}}
   	 \toprule[1.2pt]
    	 \multirow{2}{*}{PM Index} & \multicolumn{2}{c|}{Instructions} & \multicolumn{5}{c}{Last Level Cache Miss}\\
    	\cline{2-8}
	  & \clwb & \mfence & LoadA & A & B & C & E \\
   	 \midrule
   	\fastfair & 8 &10& 36 & 47 &40 & 39 & 76 \\
    	P-Bw Tree & 8 & 6 & 40  & 48 &  39 & 37& 62 \\
    	P-Masstree & 4 &  7 & 9 & 10 &8 &7 & 11 \\
    	P-ART & 3  &  4 & 4 & 5  & 5 & 5 & 22\\
    	P-HOT & 7 & 5 & 5 & 5 & 3  & 3 & 12\\
   	 \bottomrule[1.2pt]
  	\end{tabular}
}

  \mycaption{YCSB workload for tree indexes}{ The plot compares the performance of various tree based PM indexes using YCSB workloads (higher is better). All the indexes converted using \sysname outperform \fastfair, the state-of-the-art B+ tree by up to $5\times$ for string keys. For integer keys, \fastfair has better range scan performance. The fine grained performance counters per operation help explain the observed trends (lower is better).}
  \label{fig-tree}
\end{figure*}


We evaluate the performance of indexes converted using the \sysname
approach against \stateart hand-crafted PM indexes on Intel Optane DC
Persistent Memory Module (PMM).  The experiments are performed on a
2-socket, 96-core machine with 768 GB PMM, 375 GB DRAM, and 32 MB Last
Level Cache (LLC). We use the ext4-DAX file system running kernel 4.17
on the Fedora distribution. All our experiments are performed in the
\textit{App Direct} mode of Optane DC which exposes a separate
persistent memory device~\cite{appdirectmode}. All experiments are
performed on a single socket by pinning threads to a local NUMA node.
Since the machine supports \clwb instruction which is more efficient
than \clflush, we use \clwb for cache line flushes in our experiments.

We split our evaluation based on the data structure into 
ordered indexes and unordered indexes. An ordered index 
aims to support both point and range queries, but an unordered index 
only provides point queries. \fastfair, P-Bw tree, P-Masstree, P-ART, 
and P-HOT are the ordered indexes, while CCEH, Level Hashing, and 
P-CLHT are the unordered indexes. 
We use the \vtt{libvmmalloc} library from PMDK that transparently
converts traditional dynamic allocation interfaces to work on a
volatile memory pool built on a memory-mapped file on
PMEM~\cite{libvmmalloc}. We further collect low-level performance
counters such as the number of \clwb and \mfence instructions along
with the number of LLC misses per operation using the \vtt{perf} tool.

\vheading{Workloads.} We use the Yahoo! Cloud Serving Benchmark
(YCSB)~\cite{ycsb}, the industry standard for evaluating key-value
indexes. We use the index micro-benchmark to generate workload files
for YCSB and statically split them across multiple 
threads~\cite{zhang2016reducing}. Each
generated workload mimics a real application pattern as shown in
Table~\ref{tbl-ycsb}. We exclude workloads D and F as they involve
updates and some indexes (\fastfair, CCEH, CLHT) do not support key
updates. For each workload, we test two key types - \textbf{randint}
(8 byte random integer keys) and \textbf{string} (24 byte YCSB string
keys), all uniformly distributed.

To evaluate the ordered indexes, we use both random integer 
and string type keys. As the open-source implementation of \fastfair
does not support string type keys, we implement string type
support for \fastfair by replacing integer key entries with pointers
to the address of the actual string key, which is simplest way to
support variable-sized string-type keys in
B+tree in a crash-safe manner~\cite{chen2015persistent}. In both
cases, we first populate the index with 64M keys using Load A, and
then run the respective workloads that insert or read a total of 64M
keys. For unordered indexes, we only use integer key types. We present the
results from multi-threaded workloads using 16 threads and omit single
threaded results as the performance trends are comparable to the
multi-threaded workload. We use the default node size for each of the
tree-based indexes, and a starting hash table size of 48KB. The
reported numbers are averaged over several runs (with an average
variance of 0.1\%).

\subsection{Ordered indexes}

We evaluate converted indexes P-ART, P-HOT, P-Masstree, and P-BwTree
against the only concurrent and open-source \stateart PM B+ tree, \fastfair.

\vheading{Integer type keys.} P-ART outperforms \fastfair by up to
$1.6\times$ on write-heavy workloads as in Fig~\ref{fig-tree} a. The
FAST algorithm sorts inserted keys in-place, which results in higher
number of cache line flushes as compared to P-ART. This explains the
lower performance of \fastfair in write-intensive workloads.
Trie-based indexes like P-HOT eliminate key comparisons in their
search path as they do not store full keys in internal
nodes. Therefore, point reads are more cache-efficient (P-HOT incurs
$3\times$ lower LLC misses compared to \fastfair), thereby
outperforming \fastfair by $1.5\times$ on read intensive workloads.

The performance of \fastfair and P-BwTree is similar. P-BwTree
performance is low because its operations require pointer chasing; for
example, an insert can be only be performed after applying prior
deltas. This leads to many LLC misses. As a result, P-BwTree
performance is not significantly better than B+ trees with in-place
updates.

\fastfair outperforms all other indexes in range scans.  There are two
primary reasons for this. First, the keys are more compactly packed
into nodes in B+ trees unlike tries, which makes it cache efficient in
range scans. Second, the leaf nodes do not have sibling pointers in
prefix tries, thereby requiring extensive traversals for range
queries.



\vheading{String type keys.} The absolute value of throughput
decreases for string key types as compared to randint keys for all
indexes. However, the magnitude of performance drop is the highest for
\fastfair and native B+ trees, due to the high cost of string key comparison 
and pointer dereference to access the string key. This results
in $8\times$ more LLC misses in average as compared to prefix
tries. Comparing absolute throughput, we see that \fastfair performs
$2.5 - 5 \times$ worse for all YCSB workloads using string type keys
as compared to integer keys. Whereas, prefix tries are only about 20\%
slower when switched to the string keys.

As shown in fig~\ref{fig-tree}b, B+ tree's cache inefficiency results
in $3.2 - 5.2 \times$ worse performance compared to P-HOT. We observe
that although Masstree uses a data structure that is a combination of
B+ trees and prefix tries, its trie-based structure enables native key
comparison by storing 8-byte partial keys to each B+tree's
layer. Furthermore, it uses a collection of cache-friendly techniques
such as prefetching, reduced tree depth, and careful layout of data
across cachelines. These design choices makes Masstree better than its
B+ tree counterparts across all workloads.


\begin{figure}[!t]
  \centering 
{{\includegraphics[width=.45\textwidth]{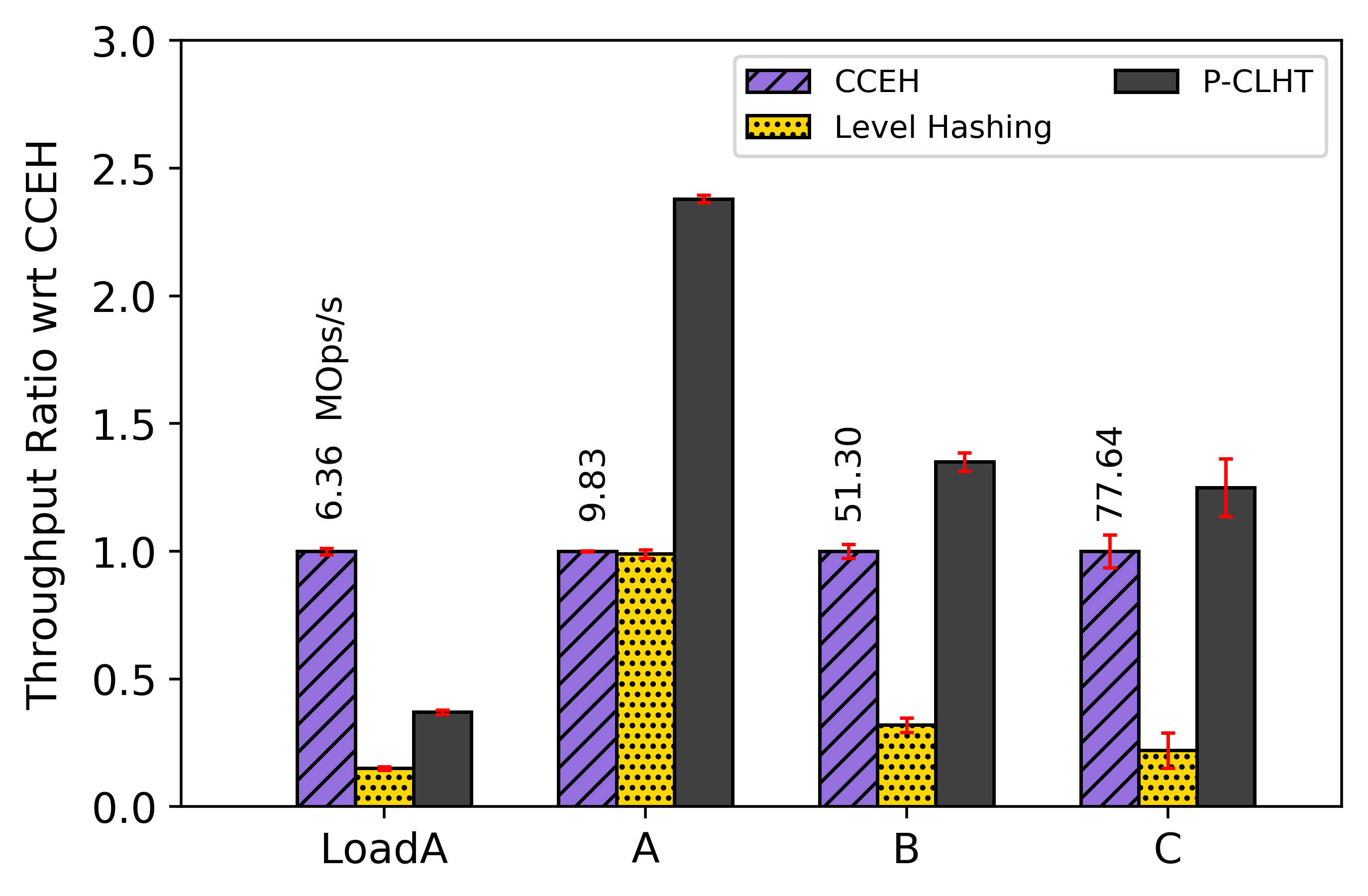} }}%
\vspace{-1em}

\mycaption{YCSB workload with integer keys for hash tables}{The figure
  compares the performance of various hash based PM indexes using YCSB
  workloads (higher is better). P-CLHT, the index converted using
  RECIPE outperforms CCEH, the state-of-the-art hash table, by up to
  $2.4\times$.}
  \label{fig-hash}
\end{figure}

\subsection{Unordered indexes}
We evaluate P-CLHT against two state-of-the-art persistent hash
tables, CCEH and Level hashing. Fig~\ref{fig-hash} shows that P-CLHT
outperforms CCEH by up to $2.5\times$ on the multi-threaded YCSB
workload. Starting from a hash table size of 48KB, we insert 64M keys
into the hash table in Load A, which triggers multiple re-hashing
operations in both indexes. P-CLHT is $2\times$ worse than CCEH for
concurrent insert only workload, due to the globally-locked rehashing
scheme that throttles concurrency. We confirm this by evaluating the
two indexes using a single thread, where P-CLHT is only 12\% slower
than CCEH even in the presence of rehashing.

Table~\ref{tbl-int-hash} shows that CCEH has lower throughput than
P-CLHT though both similar number of cache misses and \clwb
instructions. This is due to the segment split mechanism of CCEH. When
the hash table is sufficiently large, P-CLHT performs no rehashing (in
workload A and B), thereby requiring only one \clwb per insert. On the
other hand, even when similarly sized, CCEH performs frequent segment
splits that require multiple cache line flushes and expensive
copy-on-write of new segments (117K segment splits occurred on
inserting 10M keys into a sufficiently large hash table). CCEH
requires additional pointer reads due to indirections introduced using
directory and segment, which results in lower read performance over
P-CLHT.  Level hashing incurs a higher number of cache misses due to
its two level architecture that results in non-contiguous cache line
accesses~\cite{nam2019write} and lower throughput.

\subsection{Comparison to WOART}
WOART~\cite{lee2017wort} is a single-threaded, hand-crafted,
write-optimal PM variant of ART. WOART introduces a new recovery
mechanism and modifies the node structure to be failure-atomic. The
authors suggest modifying WOART to be multi-threaded using a global
lock; since this leads to low concurrency, P-ART outperforms WOART on
multi-threaded YCSB workloads by by $2 - 20 \times$.


\begin{table}[!t]
  \small
  \centering
  \ra{1.3}
  \begin{tabular}{@{}l|cc|ccccc@{}}
    \toprule[1.2pt]
     \multirow{2}{*}{PM Index} & \multicolumn{2}{c|}{Instructions} & \multicolumn{4}{c}{Last Level Cache Miss}\\
    	\cline{2-7}
	  & \clwb & \mfence & LoadA & A & B & C \\
    \midrule
    CCEH & 2.3 &  3.0 & 1.5 & 1.5 & 1.1 & 1.0    \\
    Level hashing &  3.7& 5.8 &4.0 & 3.3 &  4.0& 4.0 \\
    P-CLHT & 1.5& 2.5 & 2.4 & 1.3 & 1.1 &  1.1    \\
    \bottomrule[1.2pt]
  \end{tabular}
\vspace{0.5em} \mycaption{Performance counters}{The table shows the
  average number of \clwb, \mfence instructions per insert
  operation, and the average number of LLC misses per operation during
  each workload for randint keys (lower is better).}
  \label{tbl-int-hash}
  \vspace{-1.5em}
\end{table}

\subsection{Summary}
\sysname-converted indexes outperform \stateart hand-crafted PM
indexes by up-to $5.2 \times$ on multi-threaded YCSB
workloads. \sysname-converted indexes are optimized for cache-
efficiency and concurrency as they are built from mature DRAM
indexes. \sysname-converted indexes encounter fewer cache misses as
compare to hand-crafted PM indexes.  The append-only nature of indexes
like P-ART results in up-to $2 \times$ lower cache line flushes,
compared to hand-crafted PM indexes like \fastfair. All these factors
contribute to the performance gain of \sysname-based PM indexes.

\subsection{Testing Crash Recovery}

We test each index for 10K crash states. We load 10K entries into the
index, allowing it to crash probabilistically. We then perform a mixed
workload consisting of a total of 10K inserts and reads into the index
using 4 concurrent threads. Finally, we read back all successfully
inserted keys from the index.  On average, the end-to-end time for
generating a crash state and testing it is 20ms.

We tested the current state-of-the-art PM indexes, and our converted
PM indexes using the approach outlined in Section~\ref{sec-test}. Our
testing revealed crash-consistency bugs in \fastfair and CCEH. In
\fastfair, when two consecutive crashes occur during a node split and
a node merge, the node to be deleted by the merge algorithm is not
cleaned up correctly, which makes its right sibling inaccessible by a
reader. This results in data loss. CCEH results in stalled operations
if a crash occurs during directory doubling, as it does not update
directory metadata atomically. All PM indexes converted using \sysname
passed the testing with no bugs. Additionally, our durability test
reveals that the initial node allocation containing the root pointer
is not persisted in \fastfair and CCEH.


\section{Discussion}
\label{sec-discuss}

\vheading{Optimization}. We can increase the performance of converted
PM indexes by reducing the number of cache line flushes or memory
fences using techniques like persist buffering and
coalescing~\cite{pelley2014memory}. Persistent buffering reduces the
excessive flush and fence overhead by allowing flushes between
independent cache lines to be reordered.  Persistent coalescing
facilitates batching multiple cache line flushes to the same cache
line~\cite{cohen2017efficient, cohen2019fine}.  \sysname-based
conversion inserts a flush and fence operation after each store. We
optimized this by buffering and coalescing the flushes wherever
possible in our \sysname-converted indexes presented in
Section~\ref{sec-case}; however, such optimizations turned out to be
heavily dependent on the implementation of the index structure. As we
could not generalize these optimizations into \sklee{conditions},
\sysname leaves it to the developer to identify and apply them.

\vheading{Automation}. Converting indexes using \sklee{Condition \#1
  and \#2} only requires cache line flushes and memory fences after
every \store instruction.  Although this sounds simple and easy to
automate, the challenge in automating these conversions lies in the
many different ways in which the same logical steps are implemented in
different indexes.  For example, an atomic \store operation could be
implemented using the C++ atomic library, or through a simple pointer
assignment, followed by \mfence.

\section{Related Work}
\label{sec-related}


\vheading{Isolation and Crash Recovery}. Memory
Persistency~\cite{pelley2014memory} makes the connection between crash
recovery and the semantics of memory consistency by introducing the
concept of Recovery Observer.  \sklee{Durable
  Linearizability~\cite{izraelevitz2016linearizability} and
  Recoverable Linearizability~\cite{berryhill2016robust} theoretically
  define the relationship between crash recovery and non-blocking
  synchronization.} However, these works only propose model semantics,
without connecting the findings to practical index structures.

TSP~\cite{nawab2015procrastination} proposes the broad insight that
non-blocking indexes can be converted into crash-consistent
counterparts by coupling Recovery Observer and
Flush-on-Failure. However, Flush-on-Failure technique requires
additional hardware support like the backup power supply and kernel
modifications. \sysname, on the other hand, exploits and extends these
broad observations to build concurrent, crash-consistent PM indexes
without any hardware support and kernel changes. While TSP assumes
non-blocking writes, \sysname relaxes the assumption, allowing write
exclusion (which most concurrent DRAM indexes use).

\vheading{Concurrent Persistent Indexes}. In the past five years, 15
PM indexes have been proposed, out of which only three have open
source, concurrent implementations: \fastfair, CCEH, and Level
Hashing.
\sysname is complementary to these efforts in building a
concurrent PM index. \sysname takes a more principled approach by
reusing decades of research in building concurrent in-memory indexes
with no modifications to the underlying design of the DRAM index.

\vheading{Transactional PM Systems}. Previous work like
Atlas~\cite{chakrabarti2014atlas},
JUSTDO~\cite{izraelevitz2016failure},
NVThreads~\cite{hsu2017nvthreads}, and iDO~\cite{liu2018ido}, persist
data at boundaries of critical sections called Failure Atomic SEctions
(FASE).  They automatically inject logging for every persistent
update~\cite{chakrabarti2014atlas, hsu2017nvthreads} or program
states~\cite{izraelevitz2016failure, liu2018ido} within FASE by using
compile-time analysis. However, their approaches amplify the overhead
of cache line flushes, as they require an additional persistent
log. These systems also pay a startup cost to replay the log during
recovery, which could be significant for large indexes. However,
\sysname-converted indexes do not employ additional logging mechanisms
and pay no startup recovery cost when the index restarts after a
crash.

\vheading{Crash-Consistency Testing for PM Applications}.  PM
application testing frameworks such as Yat~\cite{yat}, Intel
PM-Inspector~\cite{pminspector}, and pmreorder~\cite{pmreorder} aim at
enabling correctness testing and debugging for applications built for
PM. However, these tools use either random or exhaustive techniques to
construct crash states, which does not scale as the number of writes
to the PM increase~\cite{yat, pminspector, pmreorder}.  Our crash
testing strategy, on the other hand, exploits the fact that operations
in PM indexes are comprised of a small set of atomic steps, thereby
simulating crashes only after these atomic steps.  This technique
makes our approach efficient and powerful enough to reveal bugs within
a few crash states.  PMTest~\cite{liu2019pmtest} requires that
developers manually annotate their source code with assert-like
statements to find errors~\cite{liu2019pmtest}. However, our approach
requires lower effort from developers, since changes are localized to
the write path.




\section{Conclusion}
\label{sec-conc}

This paper presents \sysname, a principled approach to convert
concurrent in-memory indexes to be persistent. \sysname exploits the
relationship between isolation provided by concurrent DRAM indexes and
crash recovery. \sysname provides three \sklee{conditions} to identify
DRAM indexes that can be converted to PM in a \sklee{principled}
manner, and corresponding conversion actions.  Using \sysname, we
convert five DRAM indexes, all based on different data structures to
their PM counterparts. When evaluated on Intel DC Persistent Memory,
our converted indexes outperform \stateart hand-crafted PM indexes by
as much as $5.2\times$. \sklee{\sysname-converted indexes are publicly
  available at}
\href{https://github.com/utsaslab/RECIPE}{https://github.com/utsaslab/RECIPE}.


\section*{Acknowledgments}
We would like to thank our shepherd, Peter Chen, the anonymous
reviewers, and members of Systems and Storage Lab and LASR group for
their feedback and guidance. We are also grateful to Hao Wei, Guy Blelloch,
Erez Petrank, Naama Ben-David, and Michal Friedman for their feedback on
condition \#2. This work was supported by NSF CAREER
\#1751277 and donations from VMware, Google and Facebook. We would
like to thank Intel and ETRI IITP/KEIT[2014-3-00035] for providing
access to Optane DC Persistent Memory to perform our experiments.

\newpage
\bibliographystyle{ACM-Reference-Format}
\bibliography{all}

\end{document}